\documentclass[11pt]{article}
\usepackage{latexsym,bm,graphicx,color,xcolor,nicefrac,titletoc,dirtytalk,enumerate,amsmath,amssymb,xfrac,xcolor,tensind,physics,cite,setspace}

\usepackage[nottoc]{tocbibind} 

\usepackage{hyperref}
\hypersetup{linktocpage=true,colorlinks=true,linkcolor=blue,citecolor=blue,urlcolor=blue}

\usepackage{newtxtext,newtxmath}

\DeclareMathAlphabet{\mathcal}{OMS}{cmsy}{m}{n}

\usepackage{geometry}
\usepackage{titlesec}
\titleformat*{\section}{\large\bfseries\sffamily}
\titleformat*{\subsection}{\normalsize\bfseries\sffamily}
\titleformat*{\subsubsection}{\normalsize\bfseries\sffamily}
\titleformat*{\paragraph}{\normalsize\bfseries\sffamily}

\numberwithin{equation}{section}

\newcommand{\del}{\delta}

\newcommand{\pr}{\prime}

\newcommand{\z}{\zeta}

\newcommand{\cA}{\mathcal{A}}
\newcommand{\talp}{\Tilde{\alpha}}

\newcommand{\cZ}{\mathcal{Z}}
\newcommand{\goesto}{\rightarrow}

\newcommand{\cL}{\mathcal{L}}

\newcommand{\ST}{\text{ST}}
\newcommand{\rdc}{\text{red}}
\newcommand{\onsh}{\eval_{\text{on-shell}}}
\newcommand{\IE}{I_{\text{E}}}

\newcommand{\mail}[1]{\href{mailto:#1}{{\tt #1}}}

\begin{document}
	
\begin{titlepage}       
	 \vspace{5pt} \hfill 
		
		
	\begin{center}
			\setstretch{2}
			{\Large \bf \sffamily Scaling symmetry, Smarr relation, and the extended first law in lower-dimensional Lovelock gravity}
	\end{center}
		
	\begin{center}
	\vspace{10pt}
			
	{G{\"o}khan Alka\c{c}$\,{}^{a}$, G\"{o}k\c{c}en Deniz \"Ozen$\,{}^{b}$, Hikmet \"{O}z\c{s}ahin$\,{}^{c}$, G\"{u}n S\"{u}er$\,{}^{d}$, Mustafa Tek$\,{}^{e}$}
	\\[4mm]
			
	{\small 
	{\it ${}^a$Department of Software Engineering, Faculty of Engineering and Architecture,\\ Ankara Science University, 06200 Ankara, Turkey}\\[2mm]
				
	{\it ${}^b$Department of Physics, Faculty of Science and Letters,\\Istanbul Technical University, Maslak 34469 Istanbul, Turkey}\\[2mm]
				
	{\it ${}^c$Department of Physics, Faculty of Arts and Sciences,\\ Sinop University, 57000, Sinop, Turkey}\\[2mm]
	
	{\it ${}^d$Department of Physics, Faculty of Arts and Sciences,\\ Middle East Technical University, 06800, Ankara, Turkey}\\[2mm]
	
	{\it ${}^e$Department of Physics Engineering, Faculty of Engineering and Natural Sciences,\\ Istanbul Medeniyet University, 34000, Istanbul, Turkey }\\[2mm]

	e-mail: {\mail{alkac@mail.com}, \mail{gd.ozen@gmail.com}, \mail{h.ozsahin@sinop.edu.tr}, \\ \mail{suer.gun@metu.edu.tr}, \mail{mustafa.tek@medeniyet.edu.tr}}
	}
	\vspace{5pt}
	\end{center}
		
 	\centerline{{\large \bf \sffamily{Abstract}}}
	\vspace*{5mm}
    Recently, it was discovered that lower-dimensional versions of Lovelock gravity exist as scalar-tensor theories that are examples of Horndeski gravity. We study the thermodynamics of the static black hole solutions in these theories up to cubic order through Euclidean methods. Considering solutions with spherical, planar and hyperbolic event horizons ($k=+1, 0, -1$), we show that the universality of the thermodynamics for planar black holes ($k=0$) and the extended first law that include the variation of the couplings together with their associated potentials hold also in lower dimensions. We find that in $D=4, 6$ where the 2nd- and the 3rd-order Lovelock Lagrangians are boundary terms respectively, the Smarr relation is modified since the entropy is not a homogenous function in these dimensions. We also present a derivation of the Smarr relation and its modified version based on the global scaling properties of the reduced action that is used to obtain the solutions consistently. Unlike the other hairy black hole solutions that have been analyzed before, despite the terms in the reduced action that break the scaling symmetry, the derivation still follows from a conserved Noether charge.
		
\newpage
\pagestyle{empty}
\tableofcontents
\end{titlepage}
\section{Introduction}
\setcounter{page}{1}
For a better understanding of Einstein's general relativity (GR), it is of crucial importance to find out which features are special to it and which ones can be realized in some alternatives. Among the theories that can be derived from a Lagrangian of the form $\cL[g_{\mu \nu}, R^{\mu}_{\ \nu \rho \sigma}]$, which naturally provides a generalization to Einstein tensor (symmetric and covariantly conserved) that can be consistently coupled to a covariantly conserved energy-momentum tensor, the Lovelock gravity \cite{Lanczos:1938sf,Lovelock:1971yv,Lovelock:1972vz} (see \cite{Padmanabhan:2013xyr} for a review of important properties) shares the following properties:
\begin{enumerate}[i)] %
	\item The field equations contain at most the second derivative of the metric.
	\item It possesses a unitary massless spin-2 graviton around \emph{any} of its constant curvature vacua \cite{Sisman:2012rc}.
	\item Its metric and Palatini formulations are equivalent when the torsion is zero \cite{Exirifard:2007da}. 
\end{enumerate}
The Lagrangian of Lovelock gravity is given by
\begin{equation}
	\cL = \sum_{m} c_m \cL_{m},\label{general}
\end{equation}
where the $m$-th order Lovelock Lagrangian is defined as
\begin{equation}
	\mathcal{L}_m=\frac{1}{2^m}\delta^{\mu_1\nu_1\cdots\mu_m\nu_m}_{\rho_1\sigma_1\cdots\rho_m\sigma_m}R^{\rho_1\sigma_1}_{\ \ \ \ \mu_1\nu_1}\cdots R^{\rho_m\sigma_m}_{\ \ \ \ \mu_m\nu_m}.\label{Lm}
\end{equation}
It is a boundary term when\footnote{For this reason, $D=2m$ is called the critical dimension for the $m$-th order Lagrangian $\cL_{m}$.} $D=2m$ and vanishes identically in $D<2m$. Therefore, one can have a more general theory with the above-mentioned properties only in $D>2m$.

Analogous to the Schwarzschild black hole in GR, the generic $N$-th order Lovelock gravity admits static black hole solutions with spherical, planar and hyperbolic horizons ($k=1, 0, -1$) \cite{Boulware:1985wk,Wheeler:1985nh, Wheeler:1985qd,Myers:1988ze,Cai:2003kt}. Remarkably, as later shown in \cite{o:2017sui}, the fact that it admits such solutions satisfying $g_{tt}g_{rr}=-1$ the Boyer–Lindquist coordinates is a consequence of that the theory has a unitary massless spin-2 graviton if the solution is to represent the exterior field of a spherically symmetric distribution of mass.

With or without the cosmological constant, interesting features of the thermodynamics of these solutions were discovered \cite{Myers:1988ze,Clunan:2004tb,Cai:2001dz,Cai:2003gr,Cai:2003kt,Cvetic:2001bk}. Although the equation for the metric function, which is an $N$-th order polynomial equation, and the resulting horizon structure is quite complicated, it is possible to study the thermodynamics of these solutions with relative ease by finding the relevant quantities in terms of the horizon radius \cite{Cai:2003kt} and see the effect of the Lovelock terms on the thermodynamics without the explicit form of the metric function. For example, one can show the universality of the thermodynamics of Lovelock branes ($k=0$). Although the solutions differ, their thermodynamics is identical to that of the static solution of GR. (see \cite{Cadoni:2016hhd,Hennigar:2017umz} for recent investigations).

The study of the thermodynamics of static black holes in Lovelock gravity also led to an extended first law \cite{Kastor:2010gq}, where the variations of the couplings are included. When the cosmological constant is introduced as the zeroth-order Lovelock term ($\cL_0=1$), the mass of the black hole is a function of the entropy $S$ and the coupling constants $c_{n\neq1}$, i.e., $M=M(S,c_{n\neq1})$. Since $c_{n\neq1}$ have the length unit $\ell^{2(n-1)}$, using the Euler's theorem suggest the following Smarr relation
\begin{equation}
	(D-3)M = (D-2) T S + \sum_{n\neq1} 2(n-1) \cA_n c_n,\label{smarr} 
\end{equation} 
with potentials $ \cA_n=\pdv{M}{c_n}$, and the following extended form of the first law\footnote{Throughout the paper, we derive the Smarr relation and the extended first law using the dimensionless form of the coupling constants since the equation satisfied by the metric function takes a simpler form with them.}
\begin{equation}
	\del M = T \del S + \sum_{n\neq1} \cA_n \del c_n\label{ext1st}.
\end{equation}
In \cite{Kastor:2010gq}, generalizing the findings of \cite{Kastor:2009wy} for GR with a cosmological constant, the Smarr relation \eqref{smarr} is verified geometrically from Komar integrals for the Lovelock Lagrangians \cite{Kastor:2008xb} and the extended first law \eqref{ext1st} was obtained through the Hamiltonian techniques. In this picture, a negative cosmological constant can be seen as the thermodynamics pressure and one obtains the notion of a thermodynamics volume ($\cA_0 c_0 \sim VP$), allowing the interpretation of the black hole mass $M$ as the enthalpy of the spacetime \cite{Kastor:2009wy}. In this extended phase space, novel phase transitions arise naturally (see \cite{Kubiznak:2016qmn} for a review of developments in this line of research called the black hole chemistry).

Recently, it was discovered that lower-dimensional ($D \leq 2m$) versions  of Lovelock gravity exist as scalar-tensor theories of Horndeski type \cite{Horndeski:1974wa,Kobayashi:2019hrl} with second-order field equations for the metric and the scalar field, which were obtained by two different methods that yield the same result under certain conditions. In the so-called ``Weyl trick", one evaluates the difference of the $m$-th order Lovelock Lagrangian for two conformally related metrics and finds a well-defined scalar-tensor theory after a regularization of the coupling constant $c_m$. Alternatively, one can perform a Kaluza-Klein reduction by again regularizing the relevant coupling constant with an internal space that is conformally related to a maximally symmetric space via the scalar $\phi$. In the latter approach, one obtains terms that are proportional to curvature of the maximally symmetric space. When the curvature is set to zero, the result agrees with that of the former method. Throughout this work, we will refer to these 
scalar-tensor theories as lower-dimensional Lovelock gravity. The 2nd- \cite{Fernandes:2020nbq,Lu:2020iav,Kobayashi:2020wqy,Hennigar:2020lsl} and 3rd-order \cite{Alkac:2022fuc} Lagrangians are as follows:
\begin{align}
	\cL^\ST_{2} =&\, \phi \mathcal{L}_{2}+4G^{\mu\nu}\phi_{\mu}\phi_{\nu} -4X \square\phi+2X^2, \label{ST2} \\[0.4em]
	\cL^\ST_{3}=&\, \phi \mathcal{L}_{3}-3\phi \mathcal{L}_{2}X-12R(\square\phi)^2-48R_{\mu}^{\ \alpha}R_{\nu\alpha}\phi^{\mu}\phi^{\nu} \nonumber \\ \qquad &+24R_{\mu\nu}R\phi^{\mu}\phi^{\nu}-48R^{\mu\alpha}R_{\nu\alpha\rho\mu}\phi^{\nu}\phi^{\rho}+24R_{\alpha}^{\ \sigma\mu\nu}R_{\beta\sigma\mu\nu}\phi^{\alpha}\phi^{\beta}\nonumber\\ \qquad
	&+6RX^2+24R\phi_{\mu\nu} \phi^{\mu}\phi^{\nu}+12R\phi_{\mu\nu}\phi^{\mu\nu}-96R_{\mu}^{\ \alpha} \phi^{\mu}\phi^{\nu}\phi_{\alpha\nu}\nonumber \\ \qquad
	&+48R^{\mu\nu}\square\phi\phi_{\mu\nu}+48R_{\mu\nu}\square\phi \phi^{\mu}\phi^{\nu}-48R_{\mu\nu}X\phi^{\mu}\phi^{\nu}\nonumber\\ \qquad
	&-48R^{\mu\nu}\phi_{\alpha\nu}\phi^{\alpha}_{\ \mu}-48R_{\alpha\sigma\beta\mu}\phi^{\alpha}\phi^{\beta}\phi^{\mu\sigma}-24R_{\alpha\sigma\beta\mu}\phi^{\beta\alpha}\phi^{\mu\sigma}\nonumber\\ \qquad
	&-16(\square\phi)^3+24X(\square\phi)^2+96\square\phi\phi_{\mu\nu} \phi^{\mu}\phi^{\nu}-24X^3\nonumber\\ \qquad
	&-144X\phi_{\mu\nu} \phi^{\mu}\phi^{\nu}-96 \phi^{\mu}\phi^{\nu}\phi_{\alpha\nu}\phi^{\alpha}_{\ \mu}-32\phi^{\mu\nu}\phi_{\alpha\mu}\phi^{\alpha}_{\ \nu}\nonumber \label{ST3} \\ \qquad
	&+48\square\phi\phi_{\mu\nu}\phi^{\mu\nu}-24X\phi_{\mu\nu}\phi^{\mu\nu},
\end{align}
where $\phi_{\mu}\equiv\partial_{\mu}\phi$, $\phi_{\mu\nu}\equiv \nabla_{\mu}\nabla_{\nu}\phi$ and $X\equiv\partial_{\mu}\phi\partial^{\mu}\phi$. Note that the first one is valid for $D \leq 4$ and the second one applies when $D \leq 6$.

There has been an ongoing investigation of the properties of the lower-dimensional Lovelock gravity in the literature \cite{Ma:2020ufk,Hennigar:2020fkv,Hennigar:2020drx,Khodabakhshi:2022knu,Mao:2022zrf,Alkac:2022zda,Bakopoulos:2022gdv,Guajardo:2023uix} including the thermodynamics of black hole solutions of 2nd-order theory. In this work, we aim to give an analysis of the thermodynamics of static black solutions in the up to cubic order using the Euclidean methods \cite{Regge:1974zd,Gibbons:1976ue}, which, as will be seen, makes the connection between higher-dimensional pure gravity theories and the lower-dimensional scalar-tensor theories transparent. In this approach, one gets the usual $\del M = T \del S$ form of the first law. However; after obtaining the Smarr relation from Euler's theorem and finding a modified version of it, which we will call a Smarr-like relation, in $D=4, 6$ where the entropy is not a homogeneous function (see Subsections \ref{subsec:6} and  \ref{subsec:4} for details), the extended first law for these theories can be derived easily.

Additionally, we will give a derivation of the Smarr relation and its modification that we call a Smarr-like relation from the scale symmetry of the reduced action that follows from the ansatz which can be employed to find the solution consistently. Initiated in \cite{Banados:2005hm}, this approach has been used for studying planar black holes ($k=0$) in different models \cite{Gonzalez:2009nn,Correa:2013bza,Bravo-Gaete:2014haa,Hyun:2015tia,Ahn:2015uza,Perez:2015kea,Hyun:2017nkb,Erices:2017nta,Erices:2019onl,Bravo-Gaete:2021hza} and how the method should be modified in the existence of terms breaking the scale symmetry was presented in \cite{Ahn:2015shg}. This modification provides an efficient way to obtain the holographically renormalized action and was successfully applied to various AdS/CMT models in \cite{Hyun:2016isn}. In such a case, instead of a conserved Noether charge that exists in the scale-symmetric case, one can define a charge function that evolves in the radial direction. But still, by evaluating it on the horizon and at infinity, the Smarr relation can be derived. We will show that in higher dimensions, one finds the Noether charge and then can derive the Smarr relation also for solutions with non-planar event horizons in addition to the planar case. The lower-dimensional theories that we will study in this paper exhibit an exceptional behavior that has not been observed in the literature before: Despite the scale-symmetry breaking terms in the reduced action that appear for nonplanar horizons ($k\neq0$), the charge function is conserved on-shell for the solutions under study and the Smarr(-like) relation follows straightforwardly.

The outline of this paper is as follows: In Section \ref{sec:higher}, we derive the static black hole solution of the cubic Lovelock gravity in higher dimensions, where it exists as a pure gravity theory, from a reduced action and study the thermodynamics of the solution. Then, we give a derivation of the Smarr relation based on the scale symmetry of the reduced action. In Section \ref{sec:lower}, we apply the same techniques to scalar-tensor theories that are lower-dimensional versions of cubic Lovelock gravity and discuss the similarities and differences compared to the higher-dimensional case. We end our paper with a summary and discussion of our results in Section \ref{sec:sum}.

\section{Cubic Lovelock gravity in higher dimensions}\label{sec:higher}
\subsection{General relativity with a cosmological constant in {$D=4$}}
In order to introduce the methods that we will use in a simple setup, we will first present the analysis of static black hole solutions of GR with the cosmological constant in $D=4$, whose action is given by
\begin{equation}\label{GRaction}
	I = \int \dd[4]{x} \left[ \frac{6}{L^2}\z + R  \right],
\end{equation}
which corresponds to $D=4$, $(c_0,c_1)=\left( \dfrac{6\z}{L^2},1\right)$ and $(\cL_0,\cL_1) =(1,R)$ in the general form \eqref{general}. In this paper, we will use the dimensionless form of the coupling constants in deriving the static black hole solutions and $\z$ is the first example. Instead of using the covariant form of the field equations, one can insert the ansatz
\begin{equation}\label{4dansatz}
	\dd{s}^2=-N^2(r)h(r) \dd{t}^2+\frac{\dd{r}^2}{h(r)}+r^2\dd \Sigma^{2}_{2,k},
\end{equation}
where 
\begin{equation}\label{h}
	h(r)=k + \frac{r^2}{L^2}f(r),
\end{equation}
and
\begin{equation}
	\dd \Sigma_{2,k}= \gamma_{ij} \dd x^{i} \dd x^{j}=\frac{\dd x^2}{1-kx^2}+k^2 \dd \theta^2,\qquad i,j = 2,3
\end{equation}
which corresponds to 2-dimensional spherical, planar and hyperbolic horizons ($k= +1, 0 ,-1$), into the action \eqref{GRaction} to find the following reduced action
\begin{align}
	I=& - \Sigma_2  \int \dd{t}   \int \dd{r} \frac{r}{L^2}\left[N^{\prime}\left(3 r^3 f^{\prime}+10 r^2 f+4 k L^2\right) \right. \qquad \nonumber \\
	&\left. +r N\left(r^2 f^{\prime \prime}+8 r f^{\prime}+12 f-6 \zeta\right) +2 r N^{\prime \prime}\left(r^2 f+k L^2\right)\right],
\end{align}
where $ \Sigma_2 = \int \sqrt{\gamma}\, \dd^{2} x$ and primes denote the derivative with respect to the coordinate $r$. By integrating by parts, the reduced action can be simplified into
\begin{equation}\label{GRreduced}
	I=2 \Sigma_2  \int \dd{t}   \int \dd{r} N\dv{r} \left[\frac{r^3}{L^2} \left( \zeta - f \right)\right].
\end{equation}
The equations for the functions $\left\{N(r),f(r)\right\}$ can be found from the variation of the reduced action. $\fdv{I}{N}=0$ gives the following algebraic equation for the function $f(r)$
\begin{equation}\label{algebraicf}
	\zeta-f= \frac{\omega}{r^3}, \qquad \omega:\text{constant}.
\end{equation}
An important point is that one obtains this simple form thanks to the form of the function $h(r)$ given in \eqref{h}, which will also be useful when we study more general cases. $\fdv{I}{f}=0$ is satisfied for $N(r)=1$, which determines the class of metrics that we study in this paper.

In the Euclidean approach to thermodynamics, we identify the partition function for a thermodynamic ensemble with the Euclidean path integral evaluated by the saddle-point approximation around the Euclidean continuation of the classical solution \cite{Gibbons:1976ue}. In our case, the Euclidean continuation of the metric is as follows
\begin{equation}\label{4dansatzeuc}
	\dd{s}^2_\text{E}=N^2(r)h(r) \dd{\tau}^2+\frac{\dd{r}^2}{h(r)}+r^2\dd \Sigma^{2}_{2,k},
\end{equation}
where $\tau = i t$ is the periodic Euclidean time and the Euclidean and Lorentzian actions are related by $\IE= -i I$. The periodicity of the Euclidean time is the inverse temperature of the solution which is determined by avoiding a conical singularity at the horizon. For $N(r)=1$, it reads
\begin{equation}
	\beta = \frac{1}{T} = \frac{4 \pi}{h^\pr(r_+)},\label{betadef}
\end{equation}
where $r_+$ is the location of the event horizon, which is found from the condition $h(r_+)=0$. In  terms of the metric function $f(r)$, these relations imply,
\begin{align}
	f(r_+) &= -k \frac{L^2}{r_+^2},\label{fplus}\\
	\beta &= \frac{4 \pi r_+ L^2}{-2 k L^2 + r_+^3 f^\pr(r_+)},\label{betagen}
\end{align}
which follows from eqn. \eqref{h}. Note that so far we have used only the relation between functions $h(r)$ and $f(r)$ in deriving these relations. Therefore, they will be valid when we study the generalization of the line element \eqref{4dansatz} to higher dimensions.

For static solutions of 4D GR with a cosmological constant, one has to impose the equation for the metric function $f(r)$ given in \eqref{algebraicf}. Setting $r=r_+$ in this equation (and its derivative) and using the relations (\ref{fplus}, \ref{betagen}), we find the integration constant $\omega$ and the inverse temperature $\beta$ for our solutions as follows
\begin{align}
	\omega &= L^3 \left[ \z \frac{r_+^3}{L^3} + k \frac{r_+}{L}\right],\label{wGR}\\
	\beta &= \frac{4\pi r_+}{3\z \frac{r_+^2}{L^2} + k}.\label{betaGR}
\end{align}
The Euclidean action reads
\begin{equation}\label{GReuc}
	\IE= -2 \beta \Sigma_2     \int_{r_+}^{\infty} \dd{r} N\dv{r} \left[\frac{r^3}{L^2} \left( \zeta - f \right)\right]+  \int_{r_+}^{\infty} \dd{r} \dv{B}{r},
\end{equation}
where the $\beta$ factor arises from the integration over the periodic Euclidean time and the boundary term $B$ is fixed by requiring that the Euclidean action $\IE$ has an extremum on-shell, i.e., $\var \IE\onsh =0$. For the reduced action in \eqref{GReuc}, one finds
\begin{equation}
	\var{B} =  \frac{-2 \beta \Sigma_2N r^3}{L^2}\var{f}.\label{delB}
\end{equation}
Using the equation satisfied by the metric function \eqref{algebraicf}, one sees that the on-shell Euclidean action is given by
\begin{equation}
	\IE\onsh = B\eval_{\infty} - B\eval_{r_+},\label{IEonsh}
\end{equation}
which is related to the free energy $F$ as
\begin{equation}
	\IE\onsh = \beta F = \beta M - S,
\end{equation}
where $M$ and $S$ are the black hole mass and entropy respectively. Therefore, they can be obtained from the on-shell action as follows
\begin{align}
	M &= \partial_\beta \IE\onsh,\label{Mdef}\\
	S &= \left(\beta \partial_\beta - 1\right)\IE\onsh.\label{Sdef}
\end{align}
This reduces the problem into finding the boundary term at infinity and on the horizon, for which the variation of the metric function $f(r)$ at these values of the radial coordinate is required. In studying the thermodynamics of the solutions, the variations should be taken with respect to the parameters characterizing the mass and the entropy of the solutions. In our case, we will work with the event horizon radius $r_+$.

The variation at infinity easily follows from \eqref{algebraicf} as
\begin{equation}
	\var{f}\eval_{\infty} = -\frac{\var{\omega}}{r^3}\eval_{\infty},\label{delfinf}
\end{equation}
which can be related to the changes in $r_+$ via eqn. \eqref{wGR} but we will proceed with this form because of its compactness. 

For the variation on the horizon, we should make use of the conditions $h(r_+=0)$ and $(h+\delta h)(r_+ + \delta r_+)=0$ together with the definition of the inverse temperature $\beta$ \eqref{betadef}, which imply $\eval{\delta h}_{r_+}=-\frac{4 \pi}{\beta} \delta r_+$. For the metric function, upon using \eqref{h}, this gives
\begin{equation}
	\eval{\delta f}_{r_+} = -\frac{4 \pi L^2}{\beta r_+^2} \delta r_+.\label{delfhor}
\end{equation}

The on-shell action can now be obtained by integrating the boundary term \eqref{delB} after inserting the variations (\ref{delfinf},\ref{delfhor}) as 
\begin{equation}
	\IE = \frac{2\beta\Sigma_2\omega}{L^2} - 4 \pi \Sigma_2 r_+^2,
\end{equation}
from which the mass and the entropy follows as
\begin{align}
	M &= \frac{2\Sigma_2\omega}{L^2}  =2\Sigma_2 r_+ \left[ \z \frac{r_+^2}{L^2} + k \right],\label{MGR}\\
	S &= 4 \pi \Sigma_2 r_+^2,\label{SGR}
\end{align}
where we have used \eqref{wGR} to express the mass $M$ in terms of the event horizon radius $r_+$. Note that the entropy $S$ obeys the celebrated area law.

Since we have computed $M=M(r_+,\z)$ and $S(r_+)$, using Euler's theorem we can write
\begin{align}
	M &= r_+ \pdv{M}{r_+} - 2\z \pdv{M}{\z},\label{MeulerGR}\\
	2 S &= r_+ \pdv{S}{r_+}\label{SeulerGR}.
\end{align}
From the chain rule, we have $\pdv{M}{r_+} = \frac{1}{\beta}\pdv{S}{r_+}$. Solving for $\pdv{S}{r_+}$ in \eqref{SeulerGR} and inserting the result in \eqref{MeulerGR}, we find the Smarr relation as
\begin{equation}\label{smarrGR}
	\beta M = 2 S - 2 \beta \z \cA_0, \qquad \quad \cA_0 = 2\Sigma_2 \frac{r_+^3}{L^2}.
\end{equation}
Considering the changes in the dimensionless coupling constant $\z$, the extended first law can be written as
\begin{equation}\label{extfirst}
	\beta \var M = \var S + \beta \cA_0 \var \z.
\end{equation}
This is the result first obtained in \cite{Kastor:2009wy} where the mass is interpreted as the enthalpy of the spacetime and $ \cA_0 \var \z$ term in the extended first law \eqref{extfirst} is the $V \var{P}$ term in our conventions. We will repeat the same procedure when dealing with more complicated expressions for the mass $M$ and the entropy $S$.

The Smarr relation \eqref{smarrGR} can also be derived from the global scaling symmetry of the reduced action \cite{Banados:2005hm}. If the reduced action enjoys a scaling symmetry for a Lagrangian $L_\rdc[\Phi_A,\Phi^\pr_A,r]$ which is a function of the functions $\Phi_A=\left\{N(r),f(r)\right\}$, their first derivatives $\Phi^\pr_A=\left\{N^\pr(r),f^\pr(r)\right\}$ and the coordinate $r$, the associated Noether charge is
\begin{equation}\label{chargegen}
	Q = \sum_{A} \pdv{L_\rdc}{\Phi^{'}_A} \var \Phi_{A} - r L_\rdc,
\end{equation}
where under a global scaling of the coordinate $r$ ($r \goesto \tilde{r} = \Lambda r$, $\Lambda$: constant), the scaling weights of the functions are defined as
\begin{equation}
	\Phi_{A} \goesto {\tilde{\Phi}_{A}(\tilde{r})}= \Lambda^{-\Delta_A} \Phi_{A}(r),
\end{equation}
which gives rise to the infinitesimal transformations
\begin{equation}
	\var \Phi_{A} = r \Phi_{A}^\pr + \Delta_A \Phi_{A}.
\end{equation}
Our reduced action \eqref{GReuc} is scale invariant with the scaling weights $(\Delta_N, \Delta_f)=(3, 0)$, from which the Noether charge is found to be
\begin{equation}
	Q = \frac{6 \beta \Sigma_2 N r^3 }{L^2}(\z - f),
\end{equation}
which is obviously conserved on-shell due to the equation \eqref{algebraicf} satisfied by the metric function $f(r)$ and can be written in terms of the integration constant $\omega$ as
\begin{equation}
	Q =   \frac{6 \beta\Sigma_2\omega}{L^2}.
\end{equation}
Evaluating the Noether charge $Q$ at infinity and at the horizon, we obtain
\begin{equation}
	Q\eval_{\infty}= 3 \beta M,\qquad Q\eval_{r_+}=3 \left(2 S - 2 \beta \z \cA_0\right),
\end{equation}
which are the left- and the right-hand-side of the Smarr relation \eqref{smarrGR} multiplied by 3. Hence, we have derived it from the Noether charge of the global scaling symmetry of the reduced action \eqref{GReuc}.
\subsection{Cubic Lovelock gravity in $D>6$}
Now, we are ready to analyze the thermodynamics of static black hole solutions of cubic Lovelock gravity in higher dimensions ($D>6$) where it exits as a pure gravity theory. The action of the theory is as follows
\begin{equation}\label{cubic}
	I = \int \dd[D]{x} \sqrt{-g} \left[ \frac{(D-1)(D-2)}{L^2} \zeta +  R + \alpha_2 L^2 \mathcal{L}_2 + \alpha_3 L^4 \mathcal{L}_3  \right].
\end{equation}
In the general form of the Lagrangian \eqref{general}, we have chosen
\begin{equation}
	(c_0, c_1, c_2, c_3) = \left(\frac{(D-1)(D-2)\z}{L^2}, 1, \alpha_2 L^2, \alpha_3 L^4\right),
\end{equation}
where $\z$, $\alpha_2$ and $\alpha_3$ are the dimensionless coupling constants corresponding to the 0th- ($\cL_0=1$), 2nd- and 3rd-order Lovelock Lagrangians.

The generalization of our metric ansatz to generic $D$-dimensions takes the following form
\begin{equation}\label{ansatz}
	\dd{s}^2_D = -N^2(r) h(r) \dd{t}^2 + \frac{\dd{r}^2}{h(r)} + r^2 \dd \Sigma^2_{D-2,k},
\end{equation}
where we choose the function $h(r)$ again as in \eqref{h}. The metric of $(D-2)$-dimensional constant curvature space of unit radius is given by
\begin{equation}
	d \Sigma^2_{D-2,k} =  \gamma_{ij} \dd x^{i} \dd x^{j} = \frac{\dd{x}^2}{1-kx^2} + x^2 \dd \Omega^2_{D-3},\qquad i,j = 2, \ldots D-1
\end{equation}
with $\dd \Omega^2_{D-3}= \dd \theta^2_1 + \sin^2{\theta_1} \dd \theta^2 + \dots + \sin^2\theta_1 \dots \sin^2\theta_{D-4} \dd \theta^2_{D-3}$ being the metric on a $(D-3)$-dimensional unit sphere. By appropriate identifications, it is possible to make any constant $(r, t)$ section of the spacetime compact \cite{Mann:1997iz}. Substituting this ansatz into the action and integrating by parts, we arrive at the reduced action
\begin{equation}\label{effAction2}
	I = (D-2) \Sigma_{D-2} \int \dd{t} \int \dd{r} N \dv{r} \left[ \frac{ r^{D-1}}{L^2} \left[ \zeta -  f+ \Tilde{\alpha}_2 f^2  - \Tilde{\alpha}_3 f^3  \right] \right],
\end{equation}
where $\Sigma_{D-2}= \int \sqrt{\gamma}\, \dd^{D-2} x$ and we rewrite the effect of the couplings in a compact form as
\begin{equation}\label{tildes}
	\Tilde{\alpha}_2 =(D-3)(D-4) \alpha_2, \qquad \Tilde{\alpha}_3 = (D-3)(D-4)(D-5)(D-6)\alpha_3.
\end{equation} 
We explicitly see that the 2nd- and 3rd-order Lovelock Lagrangians give a nontrivial contribution to the reduced action and the equation for the metric function $f(r)$ when $D>4$ and $D>6$ respectively. As before, $\fdv{I}{f}=0$ is satisfied for $N(r)=1$. From $\fdv{I}{N}=0$, we again find an algebraic equation for the metric function $f(r)$
\begin{equation}\label{algfcubic}
	\zeta -  f+ \Tilde{\alpha}_2 f^2  - \Tilde{\alpha}_3 f^3 = \frac{\omega}{r^{D-1}}, \qquad \omega:\text{constant}.
\end{equation}
Remember that the expressions for the metric function evaluated at the horizon radius $f(r_+)$ \eqref{fplus} and the inverse temperature \eqref{betagen} are still valid since they just follow the definition of the inverse temperature \eqref{betadef} and the relation \eqref{h} between the functions $h(r)$ and $f(r)$. Using them together with the equation \eqref{algfcubic} satisfied by the metric function $f(r)$ in cubic Lovelock gravity in higher dimensions, one finds the following expressions for the integration constant $\omega$ and the inverse temperature $\beta$
\begin{align}
	\omega &= L^2r_+^{D-3} \left[ \z \frac{r_+^2}{L^2} + k + \Tilde{\alpha}_2 k^2 \frac{L^2}{r_+^2} + \Tilde{\alpha}_3 k^3 \frac{L^4}{r_+^4} \right],\label{omegacubic} \\[0.2em]
	\beta &= 4 \pi r_+ \left[\frac{ 1 +2\Tilde{\alpha}_2 k \frac{L^2}{r_+^2} + 3\Tilde{\alpha}_3 k^2\frac{L^4}{r_+^4}}{(D-1) \z \frac{r_+^2}{L^2} + (D-3) k + (D-5) \Tilde{\alpha}_2  k^2 \frac{L^2}{r_+^2} +(D-7) \Tilde{\alpha} _3 k^3\frac{L^4}{r_+^4}} \right].\label{betacubic}
\end{align}
From the Euclidean action
\begin{equation}\label{cubiceuc}
	\IE = -(D-2) \beta \Sigma_{D-2}  \int_{r_+}^{\infty} \dd{r} N \dv{r} \left[ \frac{ r^{D-1}}{L^2} \left[ \zeta -  f+ \Tilde{\alpha}_2 f^2  - \Tilde{\alpha}_3 f^3  \right] \right] + \int_{r_+}^{\infty} \dd{r} \dv{B}{r},
\end{equation}
one can easily show that in order to ensure $\var \IE\onsh =0$, the variation of the boundary term should be taken as
\begin{equation}\label{delBcubic}
	\delta B =- \frac{(D-2) \beta \Sigma_{D-2} Nr^{D-1}}{L^2} \left[ 1 - 2 \Tilde{\alpha}_2 f+ 3 \Tilde{\alpha}_3 f^2  \right] \delta f.
\end{equation}
While the variation of the metric function at infinity can  be obtained from the algebraic eqn. \eqref{algfcubic} satisfied it, which is given by
\begin{equation}
	\eval{\delta f}_{\infty}= - \frac{\delta \omega}{\left( 1 - 2 \Tilde{\alpha}_2 f+ 3 \Tilde{\alpha}_3 f^2 \right) r^{D-1}},
\end{equation}
the variation on the horizon is again given by \eqref{delfhor} since it directly follows from the defining property of the horizon, i.e., $h(r_+)=0$, and the definition of the inverse temperature \eqref{betadef}. With these at hand, it is straightforward to find the variation of the boundary term \eqref{delBcubic} at infinity and on the horizon. From the Euclidean action \eqref{cubiceuc}, it is apparent that the on-shell action still obeys eqn. \eqref{IEonsh} with the boundary term now given by \eqref{delBcubic}. Using the relations (\ref{Mdef},\ref{Sdef}), we find the mass and the entropy of the solution as
\begin{align}
	M &= (D-2) \Sigma_{D-2} r_+^{D-3}\left[\z \frac{r_+^2}{L^2}+k+\frac{\Tilde{\alpha}_2 k^2 L^2}{r_+^2}+\frac{\Tilde{\alpha}_3 k^3 L^4}{r_+^4}\right],\\
	S &= 4 \pi \Sigma_{D-2} r_+^{D-2} \left[1+\frac{2(D-3)! \alpha_2 k L^2}{(D-1)! r_+^2} + \frac{3(D-5)! \alpha_3 k^3 L^4}{(D-1)! r_+^4} \right],\label{Sgen}
\end{align}
which agree with the results of \cite{Cai:2003kt}. Note that in the expression for the mass $M$, we use the compact form of the couplings introduced in \eqref{tildes}. In the critical dimensions where the $m$-th order Lovelock Lagrangian $\cL_{m}$ is a boundary term, the equation \eqref{algfcubic} for the metric function $f(r)$ and the mass $M$ gets no contribution. However, as explicitly seen in the eqn. \eqref{Sgen}, the entropy is modified. Also, in $D>6$ dimensions, while the equation \eqref{algfcubic} for the metric function $f(r)$ gets non-trivial contributions from the 2nd- and 3rd-order Lovelock Lagrangians $\cL_{2,3}$, for a solution with a planar horizon, the expressions for the mass and the entropy coincide with those of GR. This is the universality of the thermodynamics of Lovelock branes discussed in the literature \cite{Cadoni:2016hhd,Hennigar:2017umz}.

In order to obtain the Smarr relation, we first use Euler's theorem with $M=M(r_+,\z,\talp_2,\talp_3)$ and $S=S(r_+,\talp_2,\talp_3)$ to write
\begin{align}
	(D-3) M &= r_+ \pdv{M}{r_+}-2\z \pdv{M}{\z}+2 \talp_2 \pdv{M}{\talp_2} + 4\talp_3 \pdv{M}{\talp_3},\\
	(D-2) S &= r_+ \pdv{S}{r_+}+2 \talp_2 \pdv{S}{\talp_2} + 4\talp_3 \pdv{S}{\talp_3}.
\end{align}
Using these relations together with $\pdv{M}{r_+}= \frac{1}{\beta}\pdv{S}{r_+}$ yields the Smarr relation as follows
\begin{equation}\label{smarrcubic}
	(D-3) \beta M = (D-2) S + 2\beta \left(-\z \cA_0 + \talp_2 \cA_2+2\talp_3 \cA_3\right),
\end{equation}
where
\begin{align}
	&\begin{aligned}
		\cA_0 = (D-2) \frac{\Sigma_{D-2}r_+^{D-1}}{L^2},
	\end{aligned}\\
	&\begin{aligned}
		\cA_2 = \frac{(D-2)\Sigma_{D-2}  kL^2 r_+^{D-1}}{(D-4)\left(r_+^4 + 2 \Tilde{\alpha}_2 k L^2 r_+^2 + 3\Tilde{\alpha}_3 k^2 L^4\right)} \left[ - 2\z (D-1) \frac{r_+^2}{L^2} -(D-2)k 
		+2 \Tilde{\alpha}_2 k^2 \frac{L^2}{r_+^2} + \Tilde{\alpha}_3(D+2) k^3 \frac{L^4}{r_+^4} \right],
	\end{aligned}\\[0.2em]
	&\begin{aligned}
		\cA_3=-\frac{(D-2) \Sigma_{D-2} k^2 L^4 r_+^{D-3}}{(D-6)\left(r_+^4 + 2 \Tilde{\alpha}_2 k L^2 r_+^2 + 3\Tilde{\alpha}_3 k^2 L^4\right)} \left[3\z (D-1) \frac{r_+^2}{L^2} + (2D-3) k
		+ \Tilde{\alpha}_2(D-3) k^2 \frac{L^2}{r_+^2} -3\Tilde{\alpha}_3 k^3 \frac{L^4}{r_+^4} \right].
	\end{aligned}
\end{align}
This leads to the following form of the extended first law
\begin{equation}
	\beta \var M = \var S + \beta \left[ \cA_0 \var \z + \cA_2 \var \talp_2 + \cA_3 \var \talp_3 \right],
\end{equation}
which was first proven in \cite{Kastor:2010gq} for generic $N$-th order Lovelock gravity.

The reduced action \eqref{cubiceuc} again enjoys a scaling symmetry with the weights $(\Delta_N, \Delta_f)=(D-1, 0)$. Calculating the Noether charge from \eqref{chargegen}, we find
\begin{equation}
	Q =    \frac{(D-1)(D-2)\beta \Sigma_{D-2} N r^{D-1}}{L^2} \left[ \zeta - \Tilde{\alpha}_1 f(r) + \Tilde{\alpha}_2 f^2(r) - \Tilde{\alpha}_3 f^3(r) \right],
\end{equation}
which is conserved on-shell for $N(r)=1$ due to the equation \eqref{algfcubic} satisfied by the metric function $f(r)$. It can also be written in terms of the integration constant $\omega$ as follows
\begin{equation}
	Q =    \frac{(D-1)(D-2) \beta \Sigma_{D-2}\omega}{L^2}.
\end{equation}
When evaluated at infinity and at the horizon, it gives
\begin{equation}
	Q\eval_{\infty}= (D-1) \beta M,\qquad
	Q\eval_{r_+}=\frac{D-1}{D-3} \left[(D-2) S + 2\beta \left(-\z \cA_0 + \talp_2 \cA_2+2\talp_3 \cA_3\right)\right],
\end{equation}
which provides a derivation of the Smarr relation given in \eqref{smarrcubic}. Note that in the previous studies of Lovelock gravity \cite{Correa:2013bza,Bravo-Gaete:2021hza}, only planar black holes ($k=0$) were considered. However; we see that the scaling symmetry is present even for black holes with non-planar horizons.
\section{Cubic Lovelock gravity in lower dimensions}\label{sec:lower}
In this section, we will study the thermodynamics of the static black hole solutions of cubic Lovelock gravity in lower dimensions, which is a scalar-tensor theory. A detailed study of the solutions satisfying $g_{tt}g_{rr}=-1$ in the Boyer–Lindquist coordinates that corresponds to $N(r)=1$ in our metric ansatz \eqref{ansatz} was done in \cite{Alkac:2022fuc} via the reduced action that is obtained by employing the ansatz \eqref{ansatz} and assuming a radial profile for the scalar field, i.e., $\phi = \phi(r)$. It was found that apart from a notable exception in $D=4$, there exist scalar configurations such that the metric function $f(r)$ satisfies the algebraic equation \eqref{algfcubic} with different $\Tilde{\alpha}_i$'s ($i=2,3$) that are given in Table \ref{table}. Note that they coincide with the values obtained by scaling them by a factor of $\frac{1}{D-d}$ where $d$ is the lower dimension in which we obtain the solution and then setting $D=d$. In \cite{Glavan:2019inb}, it was claimed that this procedure defines the Lovelock gravity in lower dimensions, which was later refuted \cite{Gurses:2020ofy,Gurses:2020rxb,Arrechea:2020evj,Arrechea:2020gjw,Ai:2020peo,Fernandes:2020nbq,Mahapatra:2020rds,Bonifacio:2020vbk,Aoki:2020lig,Shu:2020cjw,Hohmann:2020cor,Cao:2021nng}. In our setting, we realize these configurations as solutions of a well-defined scalar-tensor theory.
\begin{table}[]
	\setlength{\tabcolsep}{20pt}
	\centering
	\begin{tabular}{|c|c|c|c|}
		\hline
		$D$ & $\Tilde{\alpha}_2$ & $\Tilde{\alpha}_3$ \\
		\hline
		3 & $-\alpha_2$ & $-6\alpha_3$ \\
		\hline
		4 &  $\alpha_2$ & $2\alpha_3$ \\
		\hline
		5 & $2\alpha_2$ & $-2\alpha_3$ \\
		\hline
		6 & $6\alpha_2$ & $6\alpha_3$ \\
		\hline
	\end{tabular}
	\caption{Values of $\Tilde{\alpha}_2$ and $\Tilde{\alpha}_3$ for various dimensions in terms of original dimensionless coupling constants $\alpha_2$ and $\alpha_3$}
	\label{table}
\end{table}
The scalar field configurations that give rise to these solutions (with non-zero $\Tilde{\alpha}_2$ and $\Tilde{\alpha}_3$) are given by \cite{Alkac:2022fuc}
\begin{equation}\label{phisolns}
	\phi (r)=\begin{cases}
		\log(r) \pm \int_{r_+}^r \dd{u} \frac{L}{u}\frac{|k|}{\sqrt{L^2 + k u^2 f(u)}}, & \text{$D=5,6$}\\
		\log(r). & \text{$D=3,4$}
	\end{cases}
\end{equation}
In $D=4$, the solution exists only with a planar horizon ($k=0$). When the cubic term is discarded ($\Tilde{\alpha}_3=0$), then non-planar horizons ($k=+1,-1$) are also allowed with the scalar field given by \cite{Hennigar:2020lsl,Lu:2020iav}
\begin{equation}\label{phisoln4d}
	\phi(r)= \log(r) \pm \int_{r_+}^r \dd{u} \frac{L}{u}\frac{|k|}{\sqrt{L^2 + k u^2 f(u)}}, \qquad D=4, \Tilde{\alpha}_3=0.
\end{equation}
Thermodynamical properties of these hairy black hole solutions can be found from the reduced action as we presented in the previous section. However; as will be seen, the reduced action for the scalar-tensor theories described by the Lagrangians (\ref{ST2},\ref{ST3}) does not possess a scaling symmetry. As shown in \cite{Ahn:2015shg}, one can proceed by defining a charge function $Q(r)$. If we write the Lagrangian of the reduced action $L_\rdc$ as a sum of terms with different scaling weights $L_\rdc^{[\Delta]}$ as
\begin{equation}
	L_\rdc = \sum_{\Delta} L_\rdc^{[\Delta]}, \qquad \qquad {\Tilde{L}_\rdc^{[\Delta]}} = \Lambda^{- \Delta}L_\rdc^{[\Delta]},
\end{equation}
then the charge function $Q(r)$ is defined as
\begin{equation}\label{qdef}
	Q = \sum_{A} \pdv{L_\rdc}{\Phi^{'}_A} \var \Phi_{A} - r L_\rdc, \qquad \qquad\Phi_{A} = \left\{ N, f, \xi = \phi^\pr \right\},
\end{equation}
and it satisfies
\begin{equation}\label{qder}
	\dv{Q}{r} = \sum_{\Delta} \left( \Delta - 1 \right) L_\rdc^{[\Delta]}{\onsh}.
\end{equation}
Note that when there is no term with a scaling weight other than $1$, i.e. $L_\rdc^{[\Delta\neq1]}=0$, we obtain the usual Noether charge $Q$. When the right-hand-side of the eqn. \eqref{qder} is non-zero, there is a radial evolution of the charge function $Q(r)$, which can be handled by integrating from the event horizon radius $r_+$ to infinity. We find
\begin{equation}
	Q\eval_{\infty} - Q\eval_{r_+} = \int_{r_+}^\infty \dd{r} \sum_{\Delta} \left( \Delta - 1 \right) L_\rdc^{[\Delta]}{\onsh}.
\end{equation}
By relating the charge function evaluated at infinity and at the horizon to the mass and the entropy respectively, one can still derive the Smarr(-like) relation. However; we will see that, although there are terms in the reduced action with a scaling dimension different than $1$ when the scaling weights $(\Delta_N, \Delta_f, \Delta_\xi)=(D-1, 0,1)$ that ensures the invariance for the planar black holes are assigned, they vanish on-shell for the solutions with $N(r)=1$ and $\phi(r)$ given in \eqref{phisolns} and \eqref{phisoln4d}. As a result, the charge function defined in \eqref{qdef} is still conserved. To our knowledge, these are the first examples of hairy black holes with this property. Using the conserved charge, it is straightforward to derive the Smarr(-like) relation. After the discussion of these general features, we can now move on to the details in different dimensions.
\subsection{$D=6$}\label{subsec:6}
In $D=6$, while 1st- and 2nd-order Lovelock Lagrangians $\cL_{1}=R$ and $\cL_{2}$ exist, the 3rd-order Lagrangian $\cL_{3}$ is a boundary term. Therefore, in order to have a nontrivial contribution to field equations, it should be replaced by the 3-rd order scalar-tensor version $\cL^\ST_{3}$ given in \eqref{ST3}. The action is as follows
\begin{equation}
	I = \int \dd[6]{x} \sqrt{-g} \left[ \frac{20}{L^2} \zeta +  R + \alpha_2 L^2 \mathcal{L}_2 + \alpha_3 L^4 \cL^\ST_{3} \right].
\end{equation}	
For $N(r)=1$ and the scalar field given in 	\eqref{phisolns}, the equation satisfied by the metric function $f(r)$ is the algebraic equation \eqref{algfcubic} with $\Tilde{\alpha}_{2,3}$ given in Table \ref{table}, i.e.,
\begin{equation}\label{feqn6}
	\z - f + 6 \alpha_2 f^2 - 6\alpha_3 f^3 = \frac{w}{r^5}, \qquad \omega = \text{constant}.
\end{equation}
In order to study the thermodynamics of this solution, using the metric ansatz \eqref{ansatz} in $D=6$ and assuming $\phi = \phi(r)$, we find the reduced action as
\begin{equation}
	\IE = \int_{r_+}^{\infty} \dd{r} L_{\rdc}+ \int_{r_+}^{\infty}\dd{r} \dv{B}{r}, \qquad \qquad 	L_\rdc = \sum_{\Delta} L_\rdc^{[\Delta]},
\end{equation}
where the parts of the reduced Lagrangian with different scaling weights are given by 
\begin{equation}\label{Lweight6D}
	\begin{aligned}
		&\begin{aligned}
			L_\rdc^{[1]}=\frac{4\beta \Sigma_4Nr^4}{L^2} \Big[& -6f^2\left[ 5 \alpha_2-3 \alpha_3 r^6 \xi^5 f^{\prime}+15 \alpha_3 r^5 \xi^4 f^{\prime}-30 \alpha_3 r^4 \xi^3 f^{\prime} \right. \\
			& \left. +30 \alpha_3 r^3 \xi^2 f^{\prime}-15 \alpha_3 r^2 \xi f^{\prime} \right] +f\left(5-12 \alpha_2 r f^{\prime}\right) \\
			& + r f^{\prime} + 6\alpha_3 rf^3 \left[ r^5 \xi^6+6 r^4 \xi^5+\xi^2\left(36 r^3 \xi^{\prime}-69 r\right) \right. \\
			& \left. - 6 \xi\left(4 r^2 \xi^{\prime}-5\right)+\xi^4\left(6 r^5 \xi^{\prime}-39 r^3\right) \right. \\
			& \left. +\xi^3\left(76 r^2-24 r^4 \xi^{\prime}\right)+6 r \xi^{\prime} \right] - 5 \zeta \Big],
		\end{aligned} \\[0.2em]
		&\begin{aligned}
			L_\rdc^{[3]}= \,&72\beta \Sigma_4 \alpha_3 k r^3 N(r \xi-1)^2 f \Big[ 2 r \xi\left(r^2 \xi^2-3 r \xi+2\right) f^{\prime} \\
			& +f\left(r^3 \xi^4+r \xi^2\left(6 r^2 \xi^{\prime}-19\right)+\xi\left(8-12 r^2 \xi^{\prime}\right)+6 r^2 \xi^3+2 r \xi^{\prime}\right) \Big],
		\end{aligned}\\[0.2em]
		&\begin{aligned}
			L_\rdc^{[5]}= \,&72\beta \Sigma_4 \alpha_3 k^2 L^2 r^2 N \xi(r \xi-2) \Big[ r \xi\left(r^2 \xi^2-3 r \xi+2\right) f^{\prime} \\
			&+ f\left(r^3 \xi^4+r \xi^2\left(6 r^2 \xi^{\prime}-13\right)+\xi\left(6-12 r^2 \xi^{\prime}\right)+4 r^2 \xi^3+4 r \xi^{\prime}\right) \Big],
		\end{aligned}\\[0.2em]
		&\begin{aligned}
			L_\rdc^{[7]}=24\beta \Sigma_4 \alpha_3 k^3 L^4 r N \xi^2(r \xi-2)^2\left(6 r \xi^{\prime}+r \xi^2+4 \xi\right).
		\end{aligned}
	\end{aligned}
\end{equation}
Note that we have defined a new variable $\xi = \phi^\pr$. Since the scalar field $\phi$ appears in the reduced action only with 1st- and 2nd-derivatives, we proceed with this variable such that the expression for the charge function \eqref{qdef} is still valid. Otherwise, it has to be modified for a reduced Lagrangian $L_\rdc$ that depends on the 2nd derivatives of the fields. As we see, for a general case where no choice for the function $N(r)$ is made, the global scaling symmetry is broken due to the terms with scaling weight different than 1 that arise from the 3rd-order scalar-tensor Lagrangian $\cL^\ST_{3}$.

The variation of the boundary term that follows from the condition $\var{\IE}\onsh=0$ reads
\begin{equation}
	\begin{aligned}
		\var{B} = \,&  \frac{144 \beta\Sigma_4 N r^2}{L^2} \alpha_3 (kL^2 + r^2f) \Big[ kL^2\xi(r\xi    - 2) + rf (r\xi-1)^2 \Big]^2 \var{\xi}  \\
		& + \frac{4\beta\Sigma_4 N r^3}{L^2} \Bigg[ r^2 - 12\alpha_2 r^2f + 18 \alpha_3 \xi \Big\{   k^2L^4\xi(r\xi-2)^2(r\xi-1)  \\
		& + 2kL^2rf(r\xi-2)(r\xi-1)^3 + r^3f^2\big( 5-10r\xi+10r^2\xi^2-5r^3\xi^3+r^4\xi^4 \big)    \Big\} \Bigg] \var{f},
	\end{aligned}
\end{equation}
which can be used to study the thermodynamics of a solution with $N(r) \neq 1$ if it exists. When we concentrate on the case $N(r)=1$ where the scalar field $\phi$ satisfies $\eqref{phisolns}$ and the metric function $f(r)$ satisfies \eqref{feqn6}, the variation of the boundary term becomes
\begin{equation}
	\var B = - \frac{4\beta \Sigma_4r^5}{L^2} \Big[ 1 - 12\alpha_2f + 18\alpha_3f^2 \Big] \var f.
\end{equation}
Note that there remains no contribution from the variation of the scalar field (no terms with $\var{\xi}$). Additionally, the on-shell Euclidean action still obeys the eqn. \eqref{IEonsh}. As a result, we just need to follow the same steps in the higher-dimensional case and this will be the case for all the lower-dimensional cases that we will consider in the rest of the paper. The results for the integration constant $\omega$, the inverse temperature $\beta$, the mass $M$ and the entropy $S$ are as follows
\begin{align}
	\omega &= L^2 r_+^3 \left[ \zeta \frac{r_+^2}{L^2} + k + 6\alpha_2 k^2 \frac{L^2}{r_+^2} + 6\alpha_3 k^3 \frac{L^4}{r_+^4} \right],\\[0.2em]
	\beta &= 4\pi r_+ \left[ \frac{1 + 12\alpha_2 k \frac{L^2}{r_+^2} + 18\alpha_3 k^2\frac{L^4}{r_+^4}}{5\zeta \frac{r_+^2}{L^2} + 3k + 6\alpha_2 k^2 \frac{L^2}{r_+^2} - 6\alpha_3 k^3\frac{L^4}{r_+^4}} \right],\\[0.2em]
	M &=4\Sigma_4r_+^3 \Bigg[ \z \frac{r_+^2}{L^2} + k + 6\alpha_2k^2\frac{L^2}{r_+^2} + 6\alpha_3k^3\frac{L^4}{r_+^4} \Bigg],\\[0.2em]
	S &= 4\pi\Sigma_4 r_+^4 \Bigg[ 1 + 24\alpha_2k\frac{L^2}{r_+^2} + 72\alpha_3k^2\frac{L^4}{r_+^4}\log(r_+) \Bigg].\label{S6}
\end{align}
Having obtained the mass $M=M(r_+,\z, \alpha_2, \alpha_3)$ and the entropy $S=S(r_+, \alpha_2, \alpha_3)$, one can attempt to use the Euler's theorem to obtain the Smarr relation as we did before. However; the entropy $S$ in \eqref{S6} is not a homogoneous function due to the $\log(r_+)$ contribution from the 3rd-order scalar-tensor Lagrangian  $\cL^\ST_{3}$. By a careful examination, one finds the following Smarr-like relation
\begin{equation}\label{smarr6d}
	3 \beta M = 4 S + 2 \beta \Big[ -\z \mathcal{A}_0 + \alpha_2 \mathcal{A}_2 + 2\alpha_3 \bar{\cA}_3 \Big],
\end{equation}
where
\begin{align}
	\mathcal{A}_0 &= \frac{4\Sigma_4r_+^5}{L^2},\\
	\mathcal{A}_2 &=  \frac{24\Sigma_4kL^2r_+^5}{r_+^4 + 12\alpha_2kL^2r_+^2 + 18\alpha_3k^2L^4} \Bigg[ -5\z \frac{r_+^2}{L^2} - 2k + 6\alpha_2k^2\frac{L^2}{r_+^2} + 24\alpha_3k^3\frac{L^4}{r_+^4} \Bigg], \\
	\bar{\mathcal{A}}_3 &=  \frac{6\Sigma_4k^2L^4r_+^3}{r_+^4 + 12\alpha_2kL^2r_+^2 + 18\alpha_3k^2L^4} \Bigg[  15 \z \frac{r_+^2}{L^2} \big(1 - 4 \log(r_+)\big) + k \big(13 - 36\log(r_+)\big) \nonumber \\
	& + 6\alpha_2k^2\frac{L^2}{r_+^2} \big( 11-12\log(r_+) \big) + 18\alpha_3k^3\frac{L^4}{r_+^4} \big( 3 + 4\log(r_+) \big) \Bigg].
\end{align}
While $\cA_0$ and $\cA_2$ are the potentials that appear in the extended first law, $	\bar{\mathcal{A}}_3$ is related to the potential $\cA_3$ by
\begin{equation}\label{mod}
	\bar{\cA}_3 = \cA_3 +  \frac{72\pi\Sigma_4 k^2L^4}{\beta},
\end{equation}
and the extended first law takes the following form
\begin{equation}\label{1stlaw6}
	\beta \var M = \var S +\beta \left(\cA_0 \var \z + \cA_2 \var \alpha_2 + \cA_3 \var \alpha_3\right).
\end{equation}
Since we have a modified potential $\bar{\cA}_3$ in \eqref{smarr6d} which is related to the potential $\cA_3$ in the extended first law \eqref{1stlaw6}, we prefer to call it a Smarr-like relation.

For this solution, one can calculate the charge function from \eqref{qdef}. Remarkably, it leads to the following conserved charge
\begin{equation}
	Q = \frac{20 \beta \Sigma_4 N   r^5 }{L^2} \Bigg[ \z - f + 6\alpha_2 f^2 - 6\alpha_3 f^3 \Bigg],
\end{equation}
whose derivative vanish due to \eqref{feqn6}. We have a case where the charge function $Q(r)$ does not evolve in the radial direction, and its value at infinity and at the horizon can be written as
\begin{equation}
	Q\eval_{\infty}= 5 \beta M,\qquad Q\eval_{r_+}=\frac{5}{3} \left[4 S + 2\beta \left(-\z \cA_0 + \alpha_2 \cA_2+2\alpha_3 \bar{\cA}_3\right)\right],
\end{equation}
which just implies our Smarr-like relation given in \eqref{smarr6d}.
\subsection{$D=5$}
In $D=5$, the action we consider is 	
\begin{equation}
	I = \int \dd[5]{x} \sqrt{-g} \left[ \frac{12}{L^2} \zeta +  R + \alpha_2 L^2 \mathcal{L}_2 + \alpha_3 L^4 \cL^\ST_{3} \right],
\end{equation}	
where $\cL_{3}$ in higher dimensions is replaced by the scalar-tensor Lagrangian	$\cL^\ST_{3}$ since it vanishes in $D=5$. In $\cL^\ST_{3}$ given in \eqref{ST3}, the first term vanishes and as a result of that there will be no log-type contribution to the entropy as in $D=6$. Hence, we will be able to obtain the Smarr relation in the standard form.

We study the solution with $N(r)=1$ and the scalar field $\phi$ given in \eqref{phisolns}, for which the metric function satisfies 
\begin{equation}\label{feqn5}
	\z - f + 2\alpha_2 f^2 + 2\alpha_3 f^3 = \frac{\omega}{r^4}, \qquad \omega = \text{constant}.
\end{equation}
The integration constant $\omega$ and the inverse temperature $\beta$ are
\begin{align}
	\omega &=  L^2 r_+^2 \left[ \zeta \frac{r_+^2}{L^2} + k  + 2\alpha_2 k^2 \frac{L^2}{r_+^2} - 2 \alpha_3 k^3 \frac{L^4}{r_+^4} \right], \\[3pt]
	\beta &=  2\pi r_+ \left[ \frac{1 + 4\alpha_2 k \frac{L^2}{r_+^2} - 6\alpha_3 k^2 \frac{L^4}{r_+^4}}{2\z \frac{r_+^2}{L^2} + k + 2\alpha_3k^3 \frac{L^4}{r_+^4}} \right].
\end{align}

The Lagrangian for the reduced action $L_\rdc$ can be decomposed with respect to the scaling weights as follows
\begin{equation}\label{Lweight5D}
	\begin{aligned}
		&\begin{aligned}
			L_{\rdc}^{[1]}=\frac{3\Sigma_3\beta r^3N}{L^2}\Big[& -2f^2\left[ 4 \alpha_2-12 \alpha_3 r^6 \xi^5 f^{\prime}+45 \alpha_3 r^5 \xi^4 f^{\prime} -60 \alpha_3 r^4 \xi^3 f^{\prime} \right] \\
			&+30 \alpha_3 r^3 \xi^2 f^{\prime}+4 f\left(1-\alpha_2 r f^{\prime}\right)+ r f^{\prime} \\
			& +8 \alpha_3 r^2 \xi f^3 \big[ r^4 \xi^5+\xi^2\left(34 r-18 r^3 \xi^{\prime}\right)  \\
			& +6 r^3 \xi^4+3 r^2 \xi^3\left(2 r^2 \xi^{\prime}-9\right)+3 \xi\left(6 r^2 \xi^{\prime}-5\right) - 6r\xi^{\prime} \big] - 4\zeta \Big], 
		\end{aligned}\\[0.2em]
		&\begin{aligned}
			L_{\rdc}^{[3]}&=36\Sigma_3\beta \alpha_3 k r^3 N \xi(r \xi-1) f\Big[ r \xi\left(4 r^2 \xi^2-11 r \xi+7\right) f^{\prime} \\
			& + 2 f\left[r^3 \xi^4+3 r \xi^2\left(2 r^2 \xi^{\prime}-5\right)+\xi\left(7-12 r^2 \xi^{\prime}\right)+5 r^2 \xi^3+4 r \xi^{\prime}\right] \Big], 
		\end{aligned}\\[0.2em]
		&\begin{aligned}
			L_{\rdc}^{[5]}&=18\Sigma_3\beta \alpha_3 k^2 L^2 r N \xi \Big[ r \xi\left(4 r^3 \xi^3-15 r^2 \xi^2+16 r \xi-4\right) f^{\prime} \\
			& + 4(r \xi-1) f\left(r^3 \xi^4+2 r \xi^2\left(3 r^2 \xi^{\prime}-5\right)+\xi\left(2-12 r^2 \xi^{\prime}\right)+3 r^2 \xi^3+2 r \xi^{\prime}\right) \Big], 
		\end{aligned}\\[0.2em]
		&\begin{aligned}
			L_{\rdc}^{[7]}=24\Sigma_3\beta \alpha_3 k^3 L^4 N \xi^2(r \xi-2)\Big[\xi\left(6 r^2 \xi^{\prime}-2\right)+r^2 \xi^3-6 r \xi^{\prime}+2 r \xi^2\Big].
		\end{aligned}
	\end{aligned}
\end{equation}
For a general solution, the variation of the boundary term should be taken as
\begin{equation}
	\begin{aligned}
		\var{B} = \,& \frac{144\beta\Sigma_3 N r}{L^2} \alpha_3 \xi (kL^2+r^2f) (r\xi-1)\Big[  kL^2\xi (r\xi -2) + rf(r\xi - 1)^2 \Big] \var{\xi}  \\
		& + \frac{3\beta\Sigma_3 N r^2}{L^2} \Bigg[ r^3 - 4\alpha_2fr^3 + \alpha_3 \Big\{        -24k^2L^4\xi^2 - 84kL^2r^2f\xi^2 + 96k^2L^4r\xi^3  \\
		& + 216kL^2r^3f\xi^3 - 90k^2L^4r^2\xi^4 - 180kL^2r^4\xi^4 + 24k^2L^4r^3\xi^5 +          48kL^2r^5f\xi^5  \\
		& + 6r^4f^2\xi^2 (-10+20r\xi-15r^2\xi^2+4r^3\xi^3) \Big\} \Bigg]\var{f},
	\end{aligned}
\end{equation}
which again simplifies for our solution significantly and becomes
\begin{equation}
	\var{B} = - \frac{3\beta \Sigma_3 r^4}{L^2} \Bigg[ 1 - 4\alpha_2f - 6\alpha_3f^2 \Bigg] \var{f}.
\end{equation}
From this, as explained before, it is straightforward to find the mass, the entropy and the Smarr relation as
\begin{align}
	M = & 3\Sigma_3r_+^2 \left[ \zeta \frac{r_+^2}{L^2} + k + 2\alpha_2 k^2\frac{L^2}{r_+^2} - 2\alpha_3 k^3 \frac{L^4}{r_+^4} \right], \\
	S = & 4\pi\Sigma_3 r_+^3 \left[ 1 + 12\alpha_2 k \frac{L^2}{r_+^2} + 18\alpha_3 k^2 \frac{L^4}{r_+^4} \right],\\
	2\beta M = & 3S + 2\beta \Big[ -\z \mathcal{A}_0 + \alpha_2 \mathcal{A}_2 + 2\alpha_3 \mathcal{A}_3 \Big],\label{smarr5}
\end{align}
where the potentials are given by
\begin{align}
	\mathcal{A}_0 &= \frac{3\Sigma_3r_+^4}{L^2},\\
	\mathcal{A}_2 &=   \frac{3\Sigma_3kL^2r_+^4}{r_+^4+4\alpha_2kL^2r_+^2 -6\alpha_3k^2L^4} \left[ -8\z \frac{r_+^2}{L^2} - 3k + 4\alpha_2k^2\frac{L^2}{r_+^2} - 14\alpha_3k^3\frac{L^4}{r_+^4} \right],\\
	\mathcal{A}_3 &=  \frac{6\Sigma_3k^2L^4 r_+^4}{r_+^6 + 4\alpha_2kL^2r_+^4 - 6\alpha_3k^2L^4r_+^2} \left[ 12 \z \frac{r_+^2}{L^2} + 7k + 4\alpha_2k^2\frac{L^2}{r_+^2} + 6\alpha_3k^3\frac{L^4}{r_+^4} \right],
\end{align}
and the extended first law \eqref{1stlaw6} is satisfied with these potentials.

The expression for the charge function \eqref{qdef} leads to the following Noether charge
\begin{equation}
	Q = \frac{12 \Sigma_3 \beta r^4}{L^2} \Big[ \z - f + 2\alpha_2 f^2 + 2 \alpha_3 f^3 \Big],
\end{equation}
which is conserved due to \eqref{feqn5} and gives a derivation of the Smarr relation \eqref{smarr5} with its values at infinity and the horizon given as
\begin{equation}
	Q\eval_{\infty}= 4 \beta M, \qquad Q\eval_{r_+}= 6S + 4\beta \left(-\z \cA_0 + \alpha_2 \cA_2+2\alpha_3\cA_3\right).
\end{equation}
\subsection{$D=4$}\label{subsec:4}
In $D=4$, $\cL_{3}$ vanishes and $\cL_{2}$ is a boundary term. Therefore, we replace them by the scalar-tensor Lagrangians $\cL^\ST_{2,3}$ given in (\ref{ST2}-\ref{ST3}) and consider the action
\begin{equation}
	I = \int \dd[4]{x} \sqrt{-g} \left[ \frac{6}{L^2} \zeta +  R + \alpha_2 L^2 \cL^\ST_{2} + \alpha_3 L^4 \cL^\ST_{3} \right].
\end{equation}	
As mentioned at the beginning of this section, black hole solutions with non-planar horizons exist only when $\alpha_3=0$. Before we distinguish different cases, it is useful to study the reduced action and the boundary term for a general solution described by the ansatz \eqref{ansatz} in $D=5$ and $\phi = \phi(r)$. The Lagrangian of the reduced action with different scaling weights are
\begin{equation}\label{Lweight4D}
	\begin{aligned}
		&\begin{aligned}
			L_\rdc^{[1]}=\frac{2\Sigma_2\beta r^2 N}{L^2} \bigg[& r f^2 \big[36 \alpha_3 r^5 \xi  ^5 f^{\prime}  -r^3 \xi  ^4\left(\alpha_2+90 \alpha_3 r f^{\prime}  \right) \\
			&+4 r^2 \xi^3\left(15 \alpha_3 r f^{\prime}  -\alpha_2\right) \big] +r f^2 \big[ 2 \alpha_2 r \xi  ^2\left(7-2 r^2 \xi^{\prime}  \right) \\
			& +4 \alpha_2 \xi  \left(2 r^2 \xi^{\prime}  -3\right)-4 \alpha_2 r \xi^{\prime} \big] + r f^{\prime}  \\
			& + f\left[3-2 \alpha_2 r^4 \xi^3 f^{\prime} +6 \alpha_2 r^3 \xi^2 f^{\prime} -6 \alpha_2 r^2 \xi f^{\prime} \right]  \\
			& + 6 \alpha_3 r^3 \xi  ^2 f^3\left[2 r^3 \xi^4+3 r \xi^2\left(4 r^2 \xi^{\prime}  -11\right)-4 \xi \left(6 r^2 \xi^{\prime} -5\right) \right]  \\
			& + 72 \alpha_3 f^3 r^4 \xi^2\left(r \xi^3+\xi^{\prime}\right) -3 \zeta \bigg],
		\end{aligned}\\[0.2em]
		&\begin{aligned}
			L_\rdc^{[3]} = 4\Sigma_2\beta rkN \Big[& -\alpha_2 r \xi f^{\prime}\left(r^2 \xi^2-3 r \xi +2\right) -36 \alpha_3 r^5 \xi^5 ff^{\prime}  \\
			&+2 r^2 \xi^3 f\left(\alpha_2-27 \alpha_3 r f^{\prime} \right)+\alpha_2 r \xi^2 f\left(4 r^2 \xi^{\prime} -9\right) + 2\alpha_2 r \xi^{\prime}f  \\
			& +\xi f \left(4 \alpha_2-8 \alpha_2 r^2 \xi^{\prime} \right) + 6 f^2 r^3 \alpha_3 \xi^4\left(-35+3 r^2 \xi^2+18 r^2 \xi^{\prime}\right)  \\
			& + 12 f^2 r^2 \alpha_3 \xi^2\left[9 \xi+6 r^2 \xi^3+2 r(4-9 r \xi) \xi^{\prime}\right] \\
			&+r^3 \xi^4 f\left(\alpha_2+90 \alpha_3 r f^{\prime} \right) \Big],
		\end{aligned}\\[0.2em]
		&\begin{aligned}
			L_\rdc^{[5]}=2\Sigma_2\beta L^2k^2 N \xi \Big[& 36 \alpha_3 r^3 \xi^4\left(r f^{\prime} +2 f\right) - 48 \alpha_3 r \xi^2 \left[f\left(9 r^2 \xi^{\prime}-2\right)-r f^{\prime}\right]  \\
			&+ r^2 \xi^3\left(-\alpha_2-90 \alpha_3 r f^{\prime}+6 \alpha_3 f\left(36 r^2 \xi^{\prime}-41\right)\right)  \\
			& + 36 \alpha_3 r^4 \xi^5 f+4 \xi\left(\alpha_2-r^2 \xi^{\prime}(\alpha_2-42 \alpha_3 f)\right)  +8 \alpha_2 r \xi^{\prime} \Big],
		\end{aligned}\\[0.2em]
		&\begin{aligned}
			L_\rdc^{[7]}&=24\Sigma_2\beta \alpha_3 k^3 L^4 N \xi^2\Big[ \xi^2\left(6 r^2 \xi^{\prime}-2\right)+r^2 \xi^4-12 r \xi \xi^{\prime}+4 \xi^{\prime}\Big].
		\end{aligned}
	\end{aligned}
\end{equation}
As we see, terms breaking the scaling symmetry arise from the scalar-tensor Lagrangians $\cL^\ST_{m=2,3}$ (when $k \neq 0$). The variation of the boundary term should be given by
\begin{equation}\label{varB4}
	\begin{aligned}
		\var{B} = \,& -  \frac{2\beta\Sigma_2 N}{L^2} \bigg[ \alpha_2 \Big\{ -4r^2f(kL^2+r^2f) + 8kL^2r(kL^2+r^2f)\xi  \\
		& + 8r^3f(kL^2+r^2f)\xi  - 4kL^2r^2(kL^2+r^2f)\xi^2 - 4r^4f(kL^2+r^2f)\xi^2 \Big\}  \\
		& + \alpha_3 \Big\{ 48k^2L^4(kL^2+r^2f)\xi^2  +120kL^2r^2f(kL^2+r^2f)\xi^2  \\
		& - 144k^2L^4r(kL^2+r^2f)\xi^3 - 288kL^2r^3f(kL^2+r^2f)\xi^3  \\[0.2em]
		& + 72k^2L^4r^2(kL^2+r^2f)\xi^4 + 144kL^2r^4f(kL^2+r^2f)\xi^4 \\[0.2em]
		& + 72r^4f^2(kL^2+r^2f)\xi^2(r\xi-1)^2 \Big\} \bigg] \var{\xi}  \\
		& -\frac{2N\Sigma_2\beta}{L^2} \bigg[ r^3 + \alpha_2 \Big\{-2r^2(2kL^2+3r^2f)\xi + 6r^3(kL^2+r^2f)\xi^2  \\
		& - 2r^4(kL^2+r^2f)\xi^3 \Big\} + \alpha_3 \Big\{ 48kL^2r^2(kL^2+r^2f)\xi^3 + 60r^2f(kL^2+r^2f)\xi^3  \\
		& - 90r^3(kL^2+r^2f)\xi^4 + 36r^2(kL^2r+r^3f)^2\xi^5 \Big\} \bigg]\var{f},
	\end{aligned}
\end{equation}
which can be used to study the thermodynamics of a solution.
\subsubsection{$D=4$, $\alpha_3 \neq 0$, $k=0$}
When $\alpha_3 \neq 0$, only a black hole with a planar horizon ($k=0$) is allowed with the scalar field given in \eqref{phisolns} and the metric function $f(r)$ satisfies
\begin{equation}\label{feqn4}
	\z - f + \alpha_2 f^2 - 2\alpha_3 f^3 = \frac{\omega}{r^3}, \qquad \omega = \text{constant}.
\end{equation}
The integration constant $\omega$ and the inverse temperature $\beta$ take particularly simple forms as follows
\begin{equation}
	\omega = \z r_+^3, \qquad \qquad \beta = \frac{4 \pi L^2}{3 \z r_+}.
\end{equation}
Together with the variation of the boundary term
\begin{equation}\label{delB4}
	\var B = - \frac{2 \beta \Sigma_2 r^3}{L^2} \Big[ 1 - 2\alpha_2 f + 6\alpha_3 f^2 \Big],
\end{equation}
which simplifies into $\var B = 8\pi \Sigma_2 r_+ \var r_+$, they yield the thermodynamical relations
\begin{align}
	M &= 2 \zeta\Sigma_2 \frac{r_+^3}{L^2}, \qquad S = 4\pi \Sigma_2 r_+^2, \qquad \beta M = 2S - 2\beta \z \mathcal{A}_0\label{smarr41}\\
	\beta \var{M} &= \var{S} + \beta \cA_0 \var{\z},\qquad \mathcal{A}_0 = \frac{2\Sigma_2r_+^3}{L^2}.
\end{align}
The Smarr relation in \eqref{smarr41} can be easily derived from the Noether charge
\begin{equation}\label{Q4}
	Q =  \frac{6\Sigma_2\beta r^3}{L^2} \left[ \z - f + \alpha_2 f^2 - 2\alpha_3 f^3 \right],
\end{equation}
which is conserved from \eqref{feqn4}, and whose values at infinity and the horizon are given as
\begin{equation}
	Q\eval_{\infty}= 3\beta M , \qquad Q\eval_{r_+}= 6S - 6\beta \z \cA_0.
\end{equation}
\subsubsection{$D=4$, $\alpha_3 =0$, $k=+1, 0, -1$}
For $\alpha_3 =0$, non-planar horizons are also admitted provided that the scalar field is given by \eqref{phisoln4d}. For this case, with the addition of the conformal coupling of the scalar, the mass and the entropy were computed in \cite{Babichev:2022awg} by Euclidean methods. Here, we reproduce them to discuss the scaling properties, Smarr relation and the extended first law.

The metric function satisfies the eqn. \eqref{feqn4} with $\alpha_3=0$. The integration constant $\omega$ and the inverse temperature $\beta$ are more complicated due to the $k$-dependent terms as follows
\begin{align}
	\omega &= L^2 r_+ \left[ \z \frac{r_+^2}{L^2} + k + \alpha_2 k^2 \frac{L^2}{r_+^2} \right],\\
	\beta &= 4\pi r_+ \left[ \frac{1 + 2\alpha_2 k \frac{L^2}{r_+^2} }{3\zeta \frac{r_+^2}{L^2} + k - \alpha_2 k^2 \frac{L^2}{r_+^2}} \right].
\end{align}
For this solution, although it follows from a different scalar field configuration, the variation of the boundary term is given by \eqref{delB4} with $\alpha_3=0$. Using this, the thermodynamics of the solution follows straightforwardly:
\begin{align}
	M &= 2 \Sigma_2 r_+ \left[ \z \frac{r_+^2}{L^2} + k + 2\alpha_2k^2 \frac{L^2}{r_+^2} \right], \qquad S = 4\pi \Sigma_2 r_+^2 \left[ 1 + 4\alpha_2k \frac{L^2}{r_+^2}\log(r_+)  \right], \label{S42}\\[0.2em]
	\beta M &= 2S + 2\beta \left[ -\z \mathcal{A}_0 + \alpha_2 	\mathcal{\Bar{A}}_2 \right],\qquad \beta \var{M} = \var{S} + \beta\left(\cA_0 \var{\z} + \cA_2 \var{\alpha_2}\right),\label{therm4}\\[0.4em]
	\mathcal{A}_0 &= \frac{2\Sigma_2r_+^3}{L^2}, \qquad	\mathcal{A}_2 = \frac{2\Sigma_2kL^2}{r_+} \left[ k - \frac{2(3\z r_+^4 + kL^2-\alpha_2k^2L^4)\log(r_+)}{r_+^2 + 2\alpha_2kL^2} \right],\\[0.2em]
	\mathcal{\Bar{A}}_2 &= \mathcal{A}_2 + \frac{8\pi \Sigma_2 kL^2}{\beta}. 
\end{align}
As in $D=6$ where $\cL_{3}$ is a boundary term, the entropy in \eqref{S42} is not a homogeneous function and in the Smarr-like relation in \eqref{therm4} we have a modification to the potential $\cA_2$ that appear in the extended first law in \eqref{therm4}. The Noether charge is given by the expression in \eqref{Q4} with $\alpha_3=0$ and produces the Smarr-like relation \eqref{S42} with its values at the event horizon radius $r_+$ and infinity, which are given by
\begin{equation}
	Q\eval_{\infty}= 3\beta M , \qquad Q\eval_{r_+}= 6S + 6\beta \left[ -\z \cA_0 + \alpha_2 \bar{\cA}_2 \right].
\end{equation}
\subsection{$D=3$}
Since both $\cL_{2,3}$ vanish in $D=3$, we consider the following action	\begin{equation}\label{act3}
	I = \int \dd[3]{x} \sqrt{-g} \left[ \frac{2}{L^2} \zeta +  R + \alpha_2 L^2 \cL^\ST_{2} + \alpha_3 L^4 \cL^\ST_{3} \right],
\end{equation}	
where the first term in \eqref{ST3} vanishes. 

$D=3$ is a special case for two reasons. First, since the event horizon is one-dimensional, it cannot have any curvature. Therefore, one should take $k=0$ in the metric ansatz \eqref{ansatz}. Second, in addition to the solution
\eqref{algfcubic} discussed at the beginning of this section, the static Banados-Teitelboim-Zanelli (BTZ) black hole \cite{Banados:1992wn}  is also a solution. 

We can start by considering the general form of the Lagrangian of the reduced action and the variation of the boundary term that is required to study the thermodynamics of a solution. They are given by
\begin{align}
	L_\rdc^{[1]}&=\frac{\Sigma_1\beta r}{L^2}\Big[ 2 f \left(1-2  \alpha_2 r^4 \xi ^3 f^{\prime} +3  \alpha_2 r^3 \xi ^2 f^{\prime} \right)+ r f^{\prime} - 8\alpha_2 r^3 f^2 \xi \left(\xi^2-\xi^{\prime}\right) \nonumber \\
	&+ 2 r^2 f ^2 \xi \left[36  \alpha_3 r^4 \xi ^4 f^{\prime} -r^2 \xi ^3\left( \alpha_2+45  \alpha_3 r f^{\prime} \right)+\xi \left(6  \alpha_2-4  \alpha_2 r^2 \xi^{\prime} \right)\right]  \\
	& +12  \alpha_3 r^4 f ^3 \xi ^3\left[3 \xi \left(4 r^2 \xi^{\prime} -5\right)+2 r^2 \xi ^3-12 r \xi^{\prime} +12 r \xi ^2\right]-2 \zeta\Big],  \nonumber \\[0.2em]
	\var B &=- \frac{\beta \Sigma_1 N}{L^2} \bigg[  r^2 \Big\{ 1 + 2\alpha_2 r^2 f (3-2r\xi)\xi^2 + 18\alpha_3 r^4 f^2 (4r\xi - 5)\xi^4 \Big\} \var f  \nonumber \\
	&\qquad \qquad \quad \,\,  + 8 r^4 f^2 \xi (r\xi - 1)(-\alpha_2 + 18\alpha_3 r^2 f \xi^2) \var \xi
	\bigg].\label{varB3}
\end{align}
Note that terms that break the scaling symmetry $L_\rdc^{[\Delta\neq 1] }$ in $D=6,5,4$ always come with  a $k$ factor. In $D=3$, we see that, despite the scalar hair, the Lagrangian of the reduced action has no symmetry-breaking terms. Since we need to take $k=0$ to derive the solutions, this is completely parallel to these higher dimensional cases where the event horizon might have curvature. 
\subsubsection{Solution that satisfies the polynomial equation}	For the scalar field given in \eqref{phisolns}, the metric function $f(r)$ satisfies the following polynomial equation
\begin{equation}\label{f3}
	\z - f - \alpha_2 f^2 + 6\alpha_3 f^3 = \frac{\omega}{r^2},\qquad \omega = \text{constant}.
\end{equation}
From this equation and the variation of the boundary term \eqref{varB3}, the thermodynamics of the solution can be obtained as follows:
\begin{align}
	\omega &= \z r_+^2, \qquad \qquad \beta = \frac{2 \pi L^2}{\z r_+},\\
	\var B &=- \frac{ \beta\Sigma_1 r^2}{L^2} \left( 1 + 2 \alpha_2 f - 18 \alpha_3 f^2 \right) \var f,\\
	M &= \frac{\z \Sigma_1 r_+^2}{L^2}, \qquad \qquad S = 4 \pi \Sigma_1 r_+,\\
	0 &= S + 2\beta \left[ -\z \mathcal{A}_0 + \alpha_2 \mathcal{A}_2 + 2\alpha_3 \mathcal{A}_3 \right],\label{smarr3}\\
	\mathcal{A}_0 &= \frac{\Sigma_1 r_+^2}{L^2}, \qquad \cA_2 = 0, \qquad \cA_3 = 0,\\
	\beta \var M &= \var S +\beta \left(\cA_0 \var \z + \cA_2 \var \alpha_2 + \cA_3 \var \alpha_3\right).
\end{align}
The Smarr relation \eqref{smarr3} has an alternative form in $D=3$ that is given by
\begin{equation}\label{smarralt}
	\beta M = \frac{1}{2} S,
\end{equation}
and this form of it can be derived from the Noether charge. For this solution, it reads
\begin{equation}
	Q = \frac{2\beta \Sigma_1 r^2}{L^2} \left[ \z - f - \alpha_2 f^2 + 6\alpha_3 f^3 \right],
\end{equation}
whose conservation is obvious from \eqref{f3}. Its values at infinity and at the horizon are
\begin{equation}
	Q\eval_{\infty}= 2\beta M ,\qquad Q\eval_{r_+}= S = 2\beta \z \cA_0.
\end{equation}
\subsubsection{Static BTZ black hole}
The static BTZ black hole with the line element
\begin{equation}\label{BTZ}
	\dd{s}^2 = - \left[\frac{r^2}{\ell^2} - m\right] \dd{t}^2 + \frac{1}{\dfrac{r^2}{\ell^2} - m} \dd{r}^2 + r^2 \dd{\varphi}^2,
\end{equation}
where the event horizon is located at $r_+ = \sqrt{m} \ell$, is a solution of the theory described by the action \eqref{act3} provided that the scalar field is given by \cite{Alkac:2022fuc}
\begin{equation}\label{BTZPhi}
	\phi(r) = \sqrt{\frac{\alpha_2}{18 \alpha_3}} \int_{r_+}^{r} \frac{\dd{u}}{u \sqrt{f(u)}},
\end{equation}
for which the metric function $f(r)$ becomes
\begin{equation}
	f(r) = \frac{\alpha_2^3 + 972 \z \alpha_3^2}{54\alpha_2^2 \alpha_3 + 972 \alpha_3} - \frac{m L^2}{r^2},
\end{equation}
which corresponds to the line element \eqref{BTZ} with
\begin{equation}
	\frac{\ell^2}{L^2} = \frac{54 \alpha_2^2 \alpha_3 + 972 \alpha_3^2}{\alpha_2^3 + 972 \z \alpha_3^2}.
\end{equation}
Similar to our previous computations, we can write the metric function $f(r)$ in terms of an integration constant $\omega$ as
\begin{equation}
	\z + \frac{\alpha_2^3}{972\alpha_3^2} - \left[ 1 + \frac{\alpha_2^2}{18\alpha_3} \right] f = \frac{\omega}{r^2},\qquad \omega = r_+^2 \left[ \z + \frac{\alpha_2^3}{972\alpha_3^2} \right]=\text{constant}.
\end{equation}
The inverse temperature $\beta$ in this case takes the following form
\begin{equation}
	\beta = \frac{2\pi L^2}{r_+}\left[\frac{54 \alpha_2^2 \alpha_3 + 972 \alpha_3^2}{\alpha_2^3 + 972 \z \alpha_3^2}\right].
\end{equation}
The variation of the boundary term \eqref{varB3} becomes
\begin{equation}
	\var{B} = -\frac{\beta \Sigma_1  r^2 }{L^2} \Big[ 1 + \frac{\alpha_2^2}{18\alpha_3} \Big] \var{f}= \frac{\beta\Sigma_1}{L^2} \var{\omega},
\end{equation}
and we have the following thermodynamics
\begin{align}
	M &= \frac{\Sigma_1 r_+^2}{L^2} \left[ \z + \frac{\alpha_2^3}{972\alpha_3^2} \right],\qquad S = 4\pi \Sigma_1 r_+  \Bigg[ 1 + \frac{\alpha_2^2}{18\alpha_3} \Bigg],\\
	0 &= S + 2\beta \left[ -\z \mathcal{A}_0 + \alpha_2 \mathcal{A}_2 + 2\alpha_3 \mathcal{A}_3 \right], \quad \mathcal{A}_0 = \frac{\Sigma_1r_+^2}{L^2}, \quad \cA_2 = \cA_3 = 0,\label{smarr32}\\
	\beta \var M &= \var S +\beta \left(\cA_0 \var \z + \cA_2 \var \alpha_2 + \cA_3 \var \alpha_3\right).
\end{align}
The Smarr relation in the form \eqref{smarralt} follows from the following Noether charge
\begin{equation}
	Q = \frac{\beta\Sigma_1  r^2}{486\alpha_3^2 L^2} \left[ 972\z\alpha_3^2 + \alpha_2^2 - 54\alpha_3(\alpha_2^2+18\alpha_3)f \right].
\end{equation}

For $\alpha_3=0$ in the action \eqref{act3}, the only solution for the scalar field is $\phi(r) = \text{constant}$ and one obtains the usual thermodynamics of the BTZ black hole in 3D GR with a negative cosmological constant \cite{Hennigar:2020fkv} which can be obtained by setting $\alpha_2=0$ in the thermodynamical relations above.

\section{Summary and discussions}\label{sec:sum}
In this paper, we have studied the thermodynamics of the static black hole solutions satisfying $g_{tt}g_{rr}=-1$ in lower-dimensional Lovelock gravity up to cubic term. For a comparison with the higher-dimensional case, we derived the previous results in higher dimensions where the Lovelock gravity is a pure gravity theory and shed light on a new feature. Without directly solving for the metric function, it is possible to deduce the following properties:
\begin{itemize}
	\item For planar black holes ($k=0$), the thermodynamics takes a universal form that one has the same relations as in GR \cite{Cai:2003kt}.
	\item In $D=4,6$ where the 2nd- and the 3rd-order Lovelock Lagrangians $\cL_{2,3}$ are boundary terms respectively, the metric function does not get modified due to these terms but the entropy gets a contribution from the boundary terms (when $k \neq 0$) \cite{Cai:2003kt}.
	\item By considering the variation of the couplings, one can establish an extended first law with potentials conjugate to the couplings and it is possible to obtain a Smarr relation containing these potentials \cite{Kastor:2010gq}.
	\item The reduced action that gives rise to the solutions that we consider consistently possesses a global scaling symmetry, from which it is possible to derive the Smarr relation which is compatible with the extended first law. To our knowledge, the fact that this is true for solutions with non-planar horizons ($k \neq 0$) (beyond GR) has not been noticed before.
\end{itemize}
After a detailed examination of the lower-dimensional case, we reach the following conclusions:
\begin{itemize}
	\item The universality of the thermodynamics of planar black holes still holds.
	\item In $D=4,6$, both the metric function and the entropy get modified.
	\item The extended first law still holds.
	\item The reduced action in lower dimensions is not invariant under the global scaling symmetry (when $k  \neq 0$). However; following the procedure described in \cite{Ahn:2015shg} to deal with such cases, we have shown that, for the solutions that we consider, there is a conserved Noether charge giving rise to the Smarr relation in $D=3,5$. In $D=4,6$, the contribution to the entropy from the Lagrangians $\cL_{2,3}^\ST$ comes with a $\log(r_+)$ term, which breaks the homogeneity. As a result, one obtains a Smarr-like relation which can still be obtained from the Noether charge but contains modified potentials that are related to the potentials in the extended first law.
\end{itemize}

The derivation of the Smarr(-like) relation from a conserved Noether charge despite the scaling-symmetry-breaking terms in the reduced action in lower-dimensions is quite an interesting property inherited from higher-dimensions where no such terms are present and one might wonder whether it survives when the curvature of the internal space is not set to zero. The Lagrangians up to cubic order can be found in \cite{Alkac:2022fuc}, but unfortunately, the resulting field equations are too complicated to solve in the cubic case. However; this question can be easily answered for the quadratic theory in $D=4$. It was shown in \cite{Lu:2020iav} that the modifications that arise due to the curvature of the internal space only modifies the behavior of the scalar field and the equation satisfied by the metric function remains the same. This means that when the solution for the scalar field is inserted into the reduced action, one would have
\begin{equation}
	\IE = -2 \beta \Sigma_2  \int_{r_+}^{\infty} \dd{r} N \dv{r} \left[ \frac{ r^{3}}{L^2} \left[ \zeta -  f+ \alpha_2 f^2  \right] \right] + \int_{r_+}^{\infty} \dd{r} \dv{B}{r},
\end{equation}
which is invariant with the scaling weights $(\Delta_N, \Delta_f)=(3, 0)$. Therefore, the Smarr relation still follows from the scaling symmetry. For the quadratic theory in $D=3$, a more detailed analysis is required since the modifications from the curvature of the internal space changes the solution space significantly\footnote{The solution with the metric function satisfying a quadratic equation disappears when the curvature is zero and one has only the BTZ black hole as a static black hole solution.} (see Section 2.2 of \cite{Ma:2020ufk}). We plan to study this case by considering also the rotating generalization of the BTZ black hole  together with the dual conformal field theory (CFT) interpretation in a separate work. 

We would like to emphasize that, as first noted in \cite{Alkac:2022fuc}, the Lagrangians that we study in $D=3,4,5,6$ are scalar-tensor theories with second-order field equations also in the critical dimension and beyond ($D \geq 2m$). However; when the static field configurations are studied in these dimensions, they do not admit black hole solutions, which is just another indication of that the solutions and their properties presented here are inherited from the higher-dimensional origins.

Having revealed the basic thermodynamical properties of static black hole solutions in lower-dimensional Lovelock theories, our work opens up quite a few possibilities for future work. For example, thanks to the extended phase space, the phase diagrams in the higher-dimensional versions possess quite interesting features depending on the dimension of the spacetime and the topology of the event horizon such as the appearance of the van der Waals liquid-gas system discovered first in \cite{Kubiznak:2012wp}, multiple reentrant phase transitions and tricritical points (see e.g. \cite{Cai:2013qga,Frassino:2014pha}). Our results suggest that similar results, and may be novel ones, can also be obtained in lower-dimensions.

Furthermore, since an exact black hole solution is admitted, one can also study the non-perturbative effect of the lower-dimensional Lovelock couplings on the well-known Kovtun-Son-Starinets bound on the ratio of the shear viscosity to the entropy density for the strongly coupled plasma ($\eta / s \geq \frac{1}{4 \pi}$) \cite{Kovtun:2004de}. For the quadratic theory, an analysis was given in \cite{Bravo-Gaete:2023iry}  by also considering the linear and the non-linear charged generalization of the 4D planar black hole obtained in \cite{Bravo-Gaete:2022mnr}. Apart from the non-perturbative violation of the bound which has been realized in many other different models (see references of \cite{Bravo-Gaete:2023iry}), this work is an example where one can see the efficiency of a technical tool for the calculation of the transport properties of the strongly coupled plasma described by the dual CFT that was recently developed in \cite{Fan:2018qnt}. In this method, the shear viscosity is calculated from the Noether charge corresponding to a spacelike Killing vector and it seems to be particularly useful in the analysis of the perturbations on the hairy black hole backgrounds arising from scalar-tensor theories and their charged generalizations. The planar black hole solution in the cubic theory would be a useful arena to test the efficiency of the method.

We believe that the most important applications of our results would be in the microscopic derivation of the semi-classical entropy of hairy black holes along the lines of \cite{Strominger:1997eq}, where Strominger gave a derivation of the entropy of the BTZ black hole solution of general relativity in $D=3$ with a negative cosmological constant by using the Cardy formula for the asymptotic growth of the number of states in the dual 2D CFT \cite{Cardy:1986ie}. Thanks to the investigations in \cite{Correa:2010hf,Correa:2011dt}, it is now well-established that a microscopic derivation of the entropy can be given by expressing the Cardy formula in terms of the ground state energy instead of the central charges, and identifying the negative mass soliton, which is obtained by a double Wick rotation and a redefinition of the radial coordinate in the static black hole solution, as the ground state configuration. Note that this does not contradict with the arguments of \cite{Strominger:1997eq} since the ground state there is assumed to be the thermal AdS$_3$ spacetime, which can be obtained by the same procedure. If the Newton's constant is introduced at the action level as $I = \frac{1}{16 \pi G} \int \dd[3]{x} \sqrt{-g} \, \cL$, the microscopic derivation amounts to checking the following relation
\begin{equation}
	S = 4 \pi \sqrt{-M_{\text{sol}}\, M},\label{mic3d}
\end{equation}
where $M_{\text{sol}}$ and $M$ are the mass of the soliton and the static black hole respectively. In the 3D Lovelock theory, there are two distinct hairy black hole solutions with different entropies. It would be instructive to check whether and how a microscopic derivation can be given in this case. If the formula \eqref{mic3d} is still applicable, this might imply that there are two different sectors of the dual 2D CFT characterized by different soliton configurations. To our knowledge, this would be the first such example in the literature. In principle, it might be also the case that the ground state is described by a single soliton solution describing a single ground state since, in general, integration constants disappear while redefining the radial coordinate. Therefore, an in-depth analysis should be performed.

Considering the fact that the application of the Verlinde-Cardy formula \cite{Verlinde:2000wg}, which is the higher-dimensional generalization of the Cardy formula, fails to yield the entropy of AdS black holes in higher-dimensional Lovelock gravity \cite{Cai:2001jc}, one might think that there is also no hope for the lower-dimensional Lovelock gravity other than in $D=3$. However; provided that the entropy scales as a power of the temperature as $S \sim T^\alpha$ ($\alpha$: positive constant), by considering a generalization of the modular invariance of 2D CFTs that plays a crucial role in the derivation of the Cardy formula as $\cZ[\beta] = \cZ\left[(2\pi)^{1+\alpha} \beta^{-\alpha} \right]$, the following Cardy-like formula can be obtained
\begin{equation}
	S=\frac{2(\alpha+1)\pi }{\alpha^{\frac{\alpha}{\alpha+1}}} \left(-M_{\text{sol}} \,M\right)^{\frac{1}{\alpha+1}},
\end{equation}
which reduces to \eqref{mic3d} for $\alpha = 1$, which is the usual behavior of 3D AdS black holes. For a review of the derivation of this Cardy-like formula and its extension to negative values of $\alpha$, which admits the application of it to the black holes with positive heat capacity whose most important example is the Schwarzschild black hole, we refer the reader to \cite{Hassaine:2019uyn}. Its generalization to charged, hyper-scaling violating planar black holes is also possible and it has been already successfully applied in a wide-ranging scenarious \cite{BravoGaete:2017dso}. From our results, we see that the static black hole solutions also obey the scaling condition of the entropy with $\alpha = D-2$ if the event horizon is planar ($k=0$). Therefore, we expect the higher-dimensional Cardy-formula to hold in these cases even when $D>3$, which would be a non-trivial check.

\paragraph{Note added} After we submitted our paper to arXiv, it was argued in \cite{Kubiznak:2023xbg} that the logarithmic correction to the entropy in the $D=4, \alpha_3 =0, k\neq1$ case discussed in Subsection  \ref{subsec:4}  
originate from ignoring the shift symmetry of the scalar in the field equations of the theory, which can be recovered at the action level by adding a certain boundary term. With this modification of the action, while the entropy simply obeys the area law, the first law and Smarr relation are satisfied with the same potentials but with a modified temperature that deviates from the usual relation to the surface gravity. This procedure might yield similar results for the $D=6, k\neq1$ case discussed in Subsection  \ref{subsec:6}. Additionally, the authors observe that the entropy obeys the area law in higher-dimensional Lovelock theories if the temperature is modified similar to the 4D case. In our work, the Euclidean reduced actions that we have used to study $D=4,6$ cases possess shift symmetry after integration by parts and we find the above-mentioned results with the usual relation between the temperature and the surface gravity. We refer the reader to \cite{Kubiznak:2023xbg,Liska:2023fdz} for a detailed comparison.
\paragraph*{Acknowledgements}
G. D. O. is supported by TÜBİTAK Grant No 118C587. We thank Marek Li{\v s}ka for useful correspondence on \cite{Kubiznak:2023xbg}.

\bibliographystyle{utphys}
\bibliography{refs}

\end{document}